\documentclass[conference]{IEEEtran}
\usepackage{cite}

   \usepackage{graphicx} 

\usepackage[cmex10]{amsmath}

\begin{document}
%
\title{Logic Verification of Product-Line Variant Requirements}

\author{\IEEEauthorblockN{Shamim Ripon}
\IEEEauthorblockA{Department of Computer Science and Engineering\\
East West University\\
Dhaka, Bangladesh\\
Email: dshr@ewubd.edu}
\and
\IEEEauthorblockN{Sk. Jahir Hossain, Keya Azad, Mehidee Hassan}
\IEEEauthorblockA{Department of Computer Science and Engineering\\
East West University\\
Dhaka, Bangladesh\\
Email: jahir003@gmail.com}
}

\maketitle

\begin{abstract}
Formal verification of variant requirements has gained much interest in the software product line (SPL) community. Feature diagrams are widely used to model product line variants. However, there is a lack of precisely defined formal notation for representing and verifying such models. This paper presents an approach to modeling and verifying SPL variant feature diagrams using first-order logic. It provides a precise and rigorous formal interpretation of the feature diagrams. Logical expressions can be built by modeling variants and their dependencies by using propositional connectives. These expressions can then be validated by any suitable verification tool. A case study of a Computer Aided Dispatch (CAD) system variant feature model is presented to illustrate the verification process.
\end{abstract}

\IEEEpeerreviewmaketitle

\section{Introduction}

The increase competitiveness in the software development sector with immense economic considerations such as cost, time to market, etc. motivates the transition from single product development to product-line development approach. Software product line is a set of software intensive systems sharing a common, managed set of features that satisfy the specific needs of a particular market segment or missions and that are developed from a common set of core assets in a prescribed way~\cite{Linda:SPL}.

The main idea of software product line is to explicitly identify all the requirements that are common to all members of the family as well as those that varies among products in the family. This implies a huge model that help the stakeholders to be able to trace any design choices and variability decision. A particular product is then derived by selecting the required variants and configuring them according to the product requirements.

Common requirements among all family members are easy to handle and can be integrated into the family architecture and are part of every family member. But problem arises from the variant requirements among family members. Variants are usually modeled using feature diagram, inheritance, templates and other techniques. In comparison to analysis of a single system, modeling variants adds an extra level of complexity to the domain analysis.
Different variants might have dependencies on each other. Tracing multiple occurrences of any variant and understanding their mutual dependencies are major challenges during domain modeling. While each step in modeling variants may be simple but problem arises when the volume of information grows. As a result, the impact of variant becomes ineffective on domain model. Therefore, product customization from the product line model becomes unclear and it undermines the very purpose of domain model.

This short paper presents our work-in-progress logic verification approach for variant requirements of software product line. Our particular interest is on the notion of variant dependencies that play a vital role in product customization. In our earlier work~\cite{shamim_SEN12} we have shown how a `Unified Tabular' representation along the with a decision table can be augmented with feature diagram to overcome the hurdles of variant management during an explosion of variant dependencies. However, defining such table involves manual handling of variants and hence, formal verification is not directly admissible for such approach. This paper uses first-order logic to represent product line variants and their dependencies. Such representation is amenable for various kind of formal verifications. We present a case study of Computer Aided Dispatch (CAD)\footnote{The CAD case study is adopted from Software Engineering Research group, Computer Science, National University of Singapore, http://xvcl.comp.nus.edu.sg/xvcl/cad/CAD.html} system product line  by analyzing and modeling the variants as well as the variants dependencies.

In the remainder of the paper, Section~\ref{sec:CAD} gives an overview of the CAD domain model along with a brief description of the variants and their dependencies. How variants of the CAD domain are modeled is depicted in Section~\ref{sec:Model} and a detailed feature model of CAD domain is presented as well. The formal definitions of variant models and their dependencies are presented in Section~\ref{sec:Model}. By answering various questions regarding variants models we show how the verification is performed over the logical representation of the variant model. Finally, we conclude our paper and outline our future plans in Section~\ref{sec:concl}.

\section{CAD Overview}\label{sec:CAD}

\begin{figure}[!htb]
\centering
\includegraphics[scale=.35]{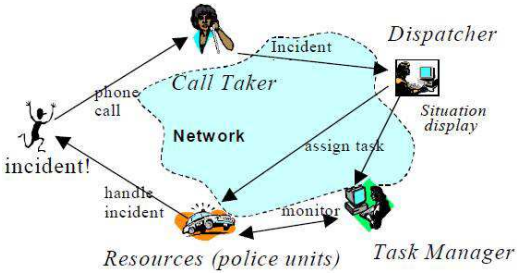}
\caption{Basic operational scenario in a CAD system for police}
\label{fig:cad}
\end{figure}

A Computer Aided Dispatch system (CAD) is a mission-critical system that is used by police, fire and rescue, health service, port operation, taxi booking and others.
Fig.~\ref{fig:cad} depicts a basic operational scenario and roles in a CAD system.

When an incident has occurred, a caller reports the incident to the command and control center of the police unit. A Call Taker in the command and control center captures the details about the incident and the Caller, and creates a task for the incident. There is a Dispatcher in the system whose task is to dispatch resources to handle any incident. The system shows the Dispatcher a list of un-dispatched tasks. The Dispatcher examines the situation, selects suitable Resources (e.g. police units) and dispatches them to execute the task.
The Task Manager monitors the situation and at the end, closes the task. Different CAD members have different resources and tasks for their system.

At the basic operational level, all CAD systems are similar; basically they support the dispatcher units to handle the incidents. However, there are differences across the CAD systems. The specific context of operation results in many variations on the basic operational theme. Some of the variants identified in CAD domain are:

\begin{itemize}
\item \emph{Call taker and dispatcher roles:} In some CAD system Call taker and dispatcher roles are separated, whereas in some system their roles are merged and one person plays the both roles.
\item \emph{Validation} of caller and task information differs across CAD systems. In some CAD systems basic validation (i.e., checking the completeness of caller information and the task information) is sufficient while in other CAD systems validation includes duplicate task checking, yet in other CAD systems no validation is required at all.
\item \emph{Un-dispatched task selection rule:} In certain situation at any given time there might be more than one task to be dispatched and it is required to decide which task will be dispatched next. A number of algorithms are available for this purpose and different CAD system use different algorithm.
\end{itemize}

This simple description of CAD variants hints us about numerous variants and their dependencies, which focus the importance of managing them properly.

\section{Modeling Variants}\label{sec:Model}

An explicit variability model as a carrier of all variability related information like specifications, interdependencies, origins, etc. can play an important and maybe the central role in successful variability management. Features are user visible aspects or characteristics of a system and are organized into And/Or graph in order to identify the commonalities and variants of the application domain. Feature modeling is an integral part of the FODA method and the Feature Oriented Domain Reuse Method (FORM)~\cite{Kang:1998}.

\begin{figure*}[!t]
\centering
\includegraphics[scale=.73]{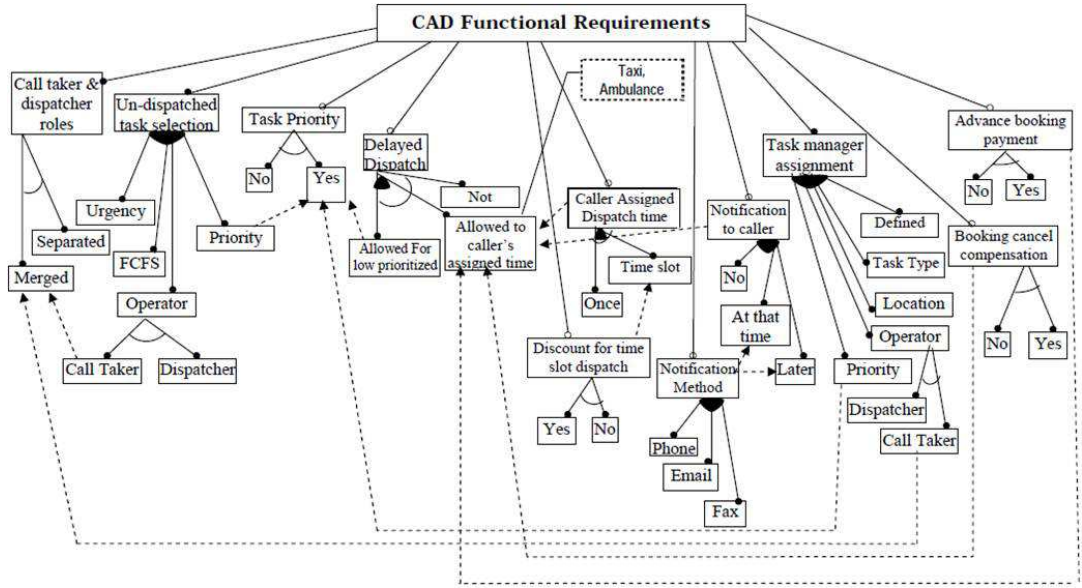}
\caption{CAD feature diagram with dependencies}
\label{fig:cadfeature}
\end{figure*}

Features are represented in graphical form as trees. The internal nodes of a tree represent the variation point and their leaves, represent the values of corresponding variation points, known as variants. Graphical symbols are used to indicate the categories of features. The root node of a feature tree always represents the domain whose features are modeled. The remaining nodes represent features which are classified into three types: \emph{Mandatory, Optional,} and \emph{Alternative}. Mandatory features are always part of the system. Optional features may be selected as a part of the system if their parent feature is in the system. The decision whether an optional feature is part of the system can be made independently from the selection of other features. Alternative features, on the other hand, are related to each other as a mutually exclusive relationship, i.e. exactly one feature out of a set of features is to be selected. There are more relationships between features. One is Or-feature~\cite{Czarnecki:2000}, which connects a set of optional features with a parent feature, either common or variant. The meaning is that whenever the parent feature is selected then at least one of the optional features will be selected. Feature diagram also depicts the interdependencies among the variants which describes the selection of one variant depends on the selection of the dependency connected variants. A CAD feature tree is illustrated in Fig.~\ref{fig:cadfeature}.

\section{Logic Representation}\label{sec:logic}

\begin{figure}[!htb]
\centering
\includegraphics[scale=.7]{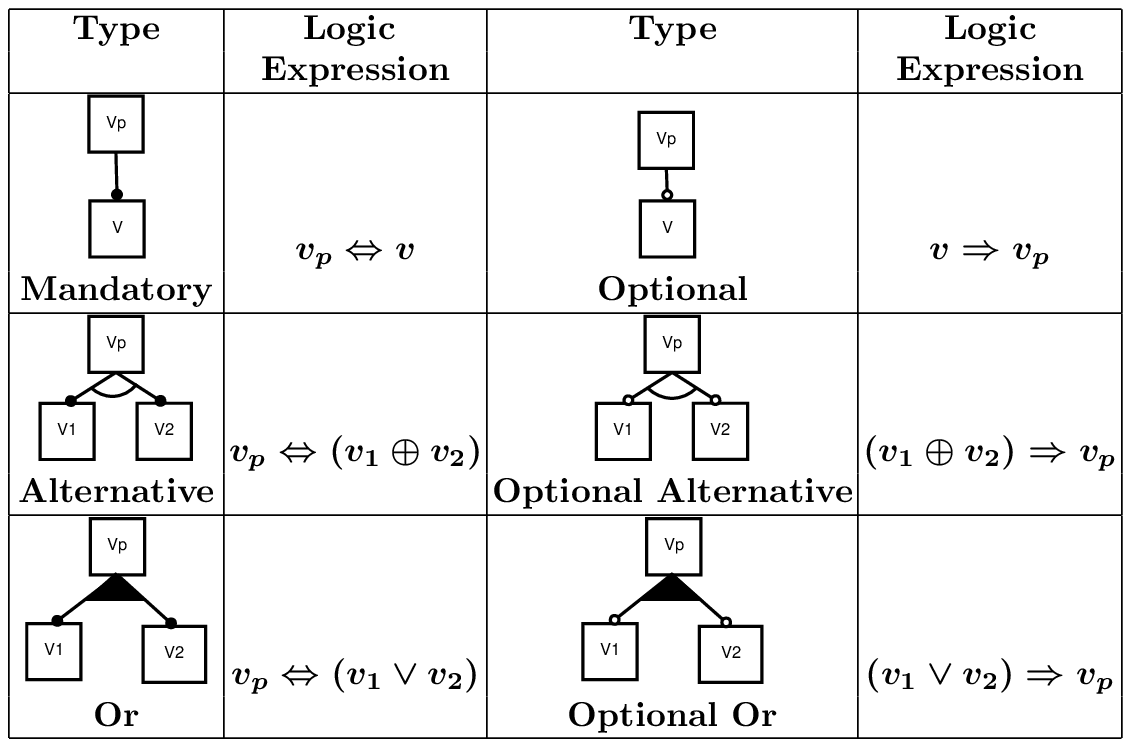}
\caption{Logical notations for feature models}
\label{fig:feature_tab}
\end{figure}

A feature model is a hierarchically arranged set of features. The relationships between a parent (or variation point) feature and its child features (variations) are categorized as follows:
\begin{itemize}
\item \emph{Mandatory}: A mandatory feature is included if its parent feature is included.
\item \emph{Optional}: An optional feature may or may not be included if its parent is included.
\item \emph{Alternative}: One and only one feature from a set of alternative features are included when parent feature is included.
\item \emph{Optional Alternative}: One feature from a set of alternative features may or may not be included if parent in included.
\item \emph{Or}: At least one from a set of \emph{or} feature is included when parent is included.
\item \emph{Optional Or}: One or more optional feature may be included if the parent is included.
\end{itemize}
The logical notations that we use in this paper to represent these features are illustrated in Fig.~\ref{fig:feature_tab}.

\begin{figure}[!htb]
\centering
\includegraphics[scale=.35]{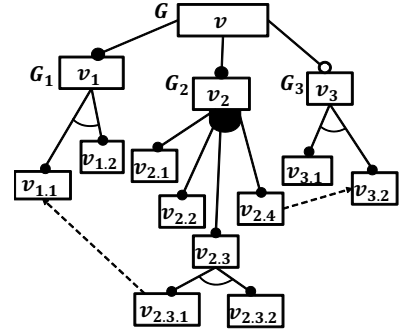}
\caption{A partial CAD feature graph using symbolic notations}
\label{fig:subgraph}
\end{figure}

A feature model (e.g. Fig.~\ref{fig:cadfeature}) can be considered as a graph consists of a set of subgraphs. Each subgraph is created separately by defining a relationship between the variation point (denoted as $v_i$) and the variants ($v_{i.j}$) by using the expressions shown in Fig.~\ref{fig:feature_tab}. For brevity, a smaller partial feature graph is drawn from CAD feature model in Fig.~\ref{fig:subgraph}. The complexity of a graph construction lies in the definition of dependencies among variants. When there is a relationship between cross-tree (or cross hierarchy) variants (or variation points) we denote it as a dependency. Typically dependencies are either \emph{inclusion} or \emph{exclusion}: if there is a dependency between $p$ and $q$, then if $p$ is included then $q$ must be included (or excluded). Only inclusion dependencies are shown in this paper. Dependencies are drawn by dotted lines. For example, there is a dotted line from $v_{2.3.1}$ to $v_{1.1}$ in Fig.~\ref{fig:subgraph}.

\subsection{Analysis of Variants}

Automatic analysis of variants are already identified as a critical task~\cite{Kang1990}. Various operations of analysis are suggested in~\cite{BenavidesCTS06, BenavidesSC10}. Our logical representation can define and validate a number of such analysis operations. The validation of a product line model is assisted by its logical representation. While constructing a single system from a product line model, we assign \texttt{TRUE (T)} value to selected variants and \texttt{FALSE (F)} to those not selected. After substituting these values to product line model, if \texttt{TRUE} value is evaluated, we call the model as valid otherwise the model is invalid. A product graph is considered to be valid if the mandatory subgraphs are evaluated to \texttt{TRUE}.

\noindent\textbf{Example 1:} Suppose the selected variants are $v_1$, $v_{1.1}$, $v_{2}$, $v_{2.1}$, $v_{2.3}$, $v_{2.3.1}$, $v_{2.4}$, $v_3$ and $v_{3.2}$. 
We check the validity of the subgraphs $G_1$, $G_2$ and $G_3$ by substituting the truth values of the variants of the subgraphs.
\begin{flalign*}
 G_1:&~~ (v_{1.1} \oplus v_{1.2}) \Leftrightarrow v_1 &\\
&= (T \oplus F) \Leftrightarrow T\\
&= T\\
G_2:&\ ~ v_2 \Leftrightarrow v_{2.1} \vee v_{2.2} \vee v_{2.3} \vee v_{2.4} &\\
&=v_2 \Leftrightarrow v_{2.1} \vee v_{2.2} \vee ((v_{2.3.1}\oplus
v_{2.3.2})\Leftrightarrow v_{2.3}) \vee v_{2.4}\\
&= T \Leftrightarrow T \vee F \vee ((T \oplus F)\Leftrightarrow T) \vee T\\
&= T\\
G_3:&~~ (v_{3.1} \oplus v_{3.2}) \Leftrightarrow v_3  &\\
&= (F \oplus T) \Leftrightarrow T \\
&= T
\end{flalign*}
As the subgraphs $G_1$, $G_2$ and $G_3$ are evaluate to \texttt{TRUE}, the product model is valid. However, variant dependencies are not considered in this case. Dependencies among variants are defined as additional constraints which must be checked separately apart from checking the validity of the subgraphs. Evaluating the dependencies of the selected variants, we get
\begin{flalign*}
\mbox{Dependency}:&~~ (v_{2.3.1} \Rightarrow v_{1.1}) \wedge (v_{2.4} \Rightarrow v_{3.2}) &\\
&= (T \Rightarrow T) \wedge (T \Rightarrow T)\\
&= T
\end{flalign*}
It concludes that the selected features from the feature model create a valid product.

\noindent\textbf{Example 2:} Similar to Example 1, suppose the selected variants are $v_1$, $v_{2}$, $v_{2.1}$, $v_{2.3}$, $v_{2.3.1}$, $v_{2.4}$, and $v_3$. Initially, neither $v_{1.1}$ nor $v_{3.2}$ is selected. However, there is inclusion dependency between $v_{2.3.1}$ and $v_{1.1}$, and between $v_{2.4}$ and $v_{3.2}$ and the dependant variants are not selected. Therefore, the whole product model becomes invalid. To handle such scenarios where dependency decision can be propagated, a set of rules has been defined using first-order logic. One of the rules indicates that if there is an inclusion dependency between $x$ and $y$ and if $x$ is selected then $y$ will be selected. Due to inclusion dependency, both $v_{1.1}$ and $v_{3.2}$ will be automatically selected and the product graph will be evaluated to \texttt{TRUE} resulting in a valid model. It indicates how the model support \emph{decision propagation}.

\noindent\textbf{Example 3:} Suppose the selected features are $v_1$, $v_{1.2}$, $v_2$, $v_{2.3}$, $v_{2.3.1}$, $v_3$ and $v_{3.1}$. For such selections each of the subgraphs is valid. Due to dependency between $v_{2.3.1}$ and $v_{1.1}$, the variant $v_{1.1}$ will be selected automatically. But $v_{1.1}$ and $v_{1.2}$ have XOR relation, hence both variants cannot be selected together. It introduces an \emph{inconsistency} into the model. It is now possible to decide which variant selection can result in an invalid model and take necessary measures.

A \emph{dead feature} is a feature that never appears in any valid product model. Identifying dead feature can optimize the product derivation from the product line model. Following the approaches shown in earlier examples, applying product model validity and decision propagation, dead features can be detected from the logical expressions. It is also possible to decide whether at least one product requirement can be selected from the product line model.

\section{Concluding Remarks}\label{sec:concl}

This short paper presented an approach to formalizing and verifying SPL variant models by using formal reasoning techniques. We provided  formal semantics of the feature models by using first-order logic and specified the definitions of six types of variant relationships. We also defined cross-tree variant dependencies. Examples are provided describing various analysis operations, such as \emph{validity, inconsistency, dead feature detection} etc. We are currently working towards answering all the analysis questions mentioned in~\cite{BenavidesCTS06, BenavidesSC10}. We are also encoding our logical notations into Prolog to be able to automatically infer any analysis operation related queries.

In contrast to other approaches \cite{Mannion:2002, ZhangICFEM, Benavides:2005:ARF, Batory05, splc/JanotaK07, gpce/CzarneckiA05}, our proposed method defines across-graph variant dependencies as well as dependencies between variation point and variants. These dependencies are defined as additional constraints while creating subgraphs from the feature graph. A knowledge-based approach to specify and verify feature models is presented in~\cite{OsmanPH08}. Comparing to that presentation, our definition relies on first-order logic which can be directly applied in many verification tools as in~\cite{Sun:2005:FSV}.

We are interested in developing an integrated variant modeling  environment that will support the construction of graphical feature models followed by the generation logical models. By using an intermediate language, such as XML, this model can then be translated to any other languages that can be directly fed into verification tools such as Alloy~\cite{Alloy}, CSP Solvers etc.





\end{document}